\documentstyle[floats,aps,epsf,prl]{revtex}

\begin{document}
\draft
\preprint{January 27, 1998}

\twocolumn[\hsize\textwidth\columnwidth\hsize\csname
@twocolumnfalse\endcsname

\title{Nature of Integer Quantum Hall Transitions} 
\author{Xiao-Qian Wang}
\address
{Department of Physics, 
Clark Atlanta University, 223 J. P. Brawley Drive, S.W., Atlanta,
Georgia 30314}
\date{Received 8 August, 2002}
\maketitle

\begin{abstract}
The
mapping between  
the metal-insulator  
transition of the quantum Hall system and a superfluid-to-insulator
transition is revisited based on a disordered anyon model.
The one-parameter scaling of the superfluid-to-insulator 
transition is employed for the analysis of the scaling behavior 
of the integer quantum Hall 
transitions.  
The analysis reveals the direct transition from a    
quantum Hall plateau to the insulator in weak disorder limit,  
and a float-down transition for strong disorders. 
In either cases, the transition corresponds to a non-chirality 
superfluid-to-insulator transition, with the longitudinal and transverse 
quantum Hall conductivities following a semi-circle relation.  
 
\end{abstract} 
\vskip1pc]

The problem of metal-insulator transition 
in quantum Hall systems has
attracted much attention [1-4] since the discovery of 
integer quantum Hall (IQH) effect. 
As the scaling theory of localization [5] predicts 
localization of all non-interacting electrons in two dimensions, a natural
challenge has been to understand the extended states at the center of
disorder-broadened Laudau bands in the IQH system. 
Laughlin [1] and Khmelnitskii [2] proposed a quasi-classical theory 
on the float-up of the extended states near the band edge.   
Based on the float-up scenario, Pruisken [3] concluded that
there exists two sets of fixed point in the 
renormalization group phase diagram for the IQH system. 
One set of stable fixed point  
is located at $\sigma^{H}_{xx}=0$ and $\sigma^{H}_{xy}=n$ 
(in the unit of $e^{2}/h$), 
corresponding to the Hall plateau. The other 
set of unstable fixed points is
located at $\sigma^{H \ast}_{xx}$ (independent of $n$) and 
$\sigma^{H \ast}_{xy}=n \pm \frac{1}{2}$,
corresponding to the center of each Laudau band,
which signals a localization-delocalization phase transition. 
A renormalization group flow diagram showing two-parameter
scaling for the IQH system was established; the Hall conductance appears as
a second coupling constant, in addition to the dissipative conductance.
The sets of fixed points are  
subsequently employed to the study of the global phase diagram of
the quantum Hall effect [4]. 

Kivelson, Lee and Zhang [4] (KLZ) proposed a global phase diagram 
for the quantum Hall system 
based on a mapping of the corresponding states of
the metal-to-insulator transition to 
ones of a superfluid-to-insulator transition. 
The theory predicted the critical conductance 
$\sigma^{H\ast}_{xx}=\frac{1}{2}$
that is supported by subsequent simulation 
studies for $n=1 \rightarrow 0$ transition [6]. 
The global phase diagram was then constructed using 
symmetry transformations of the float-up picture, 
including the particle-hole,
flux attachment, and the Laudau level 
addition transformations. By construction, the global 
phase diagram allows only nearest-neighbor IQH plateau
transitions. However, the direct transitions from high IQH plateau 
($n>1$) to  
insulator have been observed by both 
experiments [7-8] and computer simulations [9-13].  
It was argued that 
the incapability of predicting the direct transitions in the global phase
diagram is attributed  
to the  
lattice effects that are not present in the continuum models. 

In this Letter, I revisit the mapping between the metal-to-insulator 
transitions and the superfluid-to-insulator transition based on a 
disordered anyon model [14].
The resultant 
mapping relations are essentially the same as those obtained by KLZ in the
weak disorder limit. However, closer scrutiny of the
self-duality relations without invoking 
the symmetry transformations {\it a priori}, reveals the existence of 
direct transitions form high quantum Hall plateaus to the insulator. 
This finding and the associated predictions based
on one-parameter scaling flow are remarkably in agreement
with the numerical simulations [9-13]; thereby providing important
information on the microscopic
origin of the direct transitions. Subsequently, we consider the strong
disorder limit where the  
float-down transitions 
become appropriate.                
A distinct feature of our analysis is the realization that 
a superfluid-to-insulator transition with non-chirality constraint, 
maps to a quantum Hall critical region 
where the longitudinal and transverse quantum Hall conductivities   
satisfy a semi-circle relation. The semi-circle relation was
proposed by Ruzin and co-workers [15] based on a 
phenomenological two-fluid theory.   

$Statistical$ $mapping.-$The disordered 
anyon model is described by the Hamiltonian [14,16]
\begin{equation}
\hat{H} = \sum_i \left\{ \frac{1}{2m} \left(
{\bf P}_i - \hat{\bf A}({\bf r}_{i})
\right)^{2}
+ V({\bf r}_{i}) \right\} ,
\end{equation}
where $V({\bf r}_{i})$ is short-range correlated random potential,
and $\hat{\bf A}({\bf r}_{i})$
is a
statistical field operator given by
\begin{equation}
\hat{\bf A}({\bf r}_{i})\equiv \frac{\hbar}{n} \sum_j 
\frac{\hat {\bf z}\times
({\bf r}_{i}-{\bf r}
_j)}{| {\bf r}_{i}-{\bf r}_j |^2},
\end{equation}
with $\hat {\bf z}$ a unit vector normal to the
plane,
and 
$n$ the so-called statistics     
parameter ($n=1$ for bosons; $n=2$ for semions; and $n=\infty$ for fermions)
characterizing the specific form of the fractional
statistics. The system is equivalent to spinless fermions interacting
through Chern-Simons gauge field and disordered impurity scattering. 
It is worth noting that disordered-fermion system
stands for the $n=\infty$ limit of the model.

We
first consider
the average statistical field that yields
a fictitious magnetic field ${\bf B}=B \hat{\bf z}$. The
resulting mean-field Hamiltonian 
\begin{equation}
\hat{H_{0}} = \sum_i \frac{1}{2m} ({\bf P}_i 
- \bar{\bf A}_i)^2+ V({\bf r}_{i}),
\end{equation}
describes a collection of 
spinless particles moving in a strong magnetic field ${\bf B}$ with
$n$ filled Laudau bands and the 
random potential $V$. 
The equivalence between the IQH system and the 
mean-field solution of the disordered anyon system is 
crucial in the statistical mapping.

The effect of statistical-field fluctuations around 
the mean-field solution, $\hat{H} -\hat{H_{0}} $, 
can be evaluated by a consistent  
$1/n$-expansion scheme [17]. In the absence of the random potential, 
the leading-$1/n$-order
results are the same as those obtained
in the calculation of the random-phase-approximation (RPA) response
functions [14,16]. 
Based on a correlated RPA calculation [17],
we have shown that the next-leading-order corrections to the
RPA results have no effect in the long-wavelength limit. 
As a consequence, the 
longitudinal and transverse conductivities, $\sigma_{xx}$
and $\sigma_{xy}$,
can be evaluated from  
the leading-$1/n$-order current-current correlation functions at $T=0$,
and are related to those for IQH systems, 
$\sigma_{xx}^{H}$ and $\sigma_{xy}^{H}$,  by expressions [14,17]: 
\begin{eqnarray}
\sigma_{xx}&=&n^{2}\frac{\sigma_{xx}^{H}}{(\sigma_{xx}^{H})^{2}
+ (\sigma_{xy}^{H}-n)^{2}},  \\
\sigma_{xy}&=&\sigma_{R}+n^{2}\frac{\sigma_{xy}^{H}-n}{(\sigma_{xx}^{H})^{2}
+ (\sigma_{xy}^{H}-n)^{2}}.
\end{eqnarray}
Here $\sigma_{R}=n$ in the absence of disorders. This set of equations
is equivalent to that obtained by KLZ [4] via a transformation $n \rightarrow
1/n$.
In fact, they become identical if one changes the unit
of conductivities in Eqs. (4-5) 
from $e^{2}/h$ to $e^{\ast 2}/h$ where $e^{\ast}=ne$. 
 
The mapping of conductivities between the disordered anyon system and
the IQH system is self-dual in that exchanging 
$(\sigma_{xx}^{H},\sigma_{xy}^{H})$ with $(\sigma_{xx},\sigma_{xy})$,
the function form of Eqs. (4-5) remains invariant. The self-duality
is the ramification
of the particle-vortex duality in connection to the
Chern-Simons gauge field.  
 
It is
worth pointing out that the effect of disorder plays an important
role in the IQH systems. As such, it is necessary to evaluate the
effect of disorder on the mapping relations. In the weak disorder limit,
the self-duality equations was argued to be intact [14]. 
The weak-disorder limit 
refers to the case that there exists IQH effect for $n$ Landau bands. 
In this case, the contribution to $\sigma_{R}$ for the anyon model
can be calculated
using the quasi-classical theory [1-2].
  
Considerable insight 
of the scaling behavior of the quantum Hall transitions can
be gained from the mapping [4,14].
To this end, we first construct the corresponding states 
between the two systems. From Eqs. (4-5),  
$(\sigma_{xx}^{H},\sigma_{xy}^{H})=(0,0)$ maps to 
$(\sigma_{xx},\sigma_{xy})=(0,0)$, corresponding to 
an insulating state.  
The ``localization'' fixed points associated with
the plateau solutions in the IQH system correspond to
two types of states in the disordered anyon system.
$(\sigma_{xx}^{H},\sigma_{xy}^{H})=(0,m)$ ($m \neq n$) maps to
a quantum Hall conductor, and $(\sigma_{xx}^{H},\sigma_{xy}^{H})=(0,n)$  
was shown to map to $(\sigma_{xx},\sigma_{xy})=(\infty,n)$, corresponding to
an anyon-type superfluid state
[14],
characterized by 
a spontaneous breaking of parity and time-reversal invariance.
Based on the concept of corresponding states, KLZ [4] constructed the 
global phase diagram using symmetry transformations.  

{\it Direct transition.}$-$While the corresponding 
state for the insulator is unique,
the corresponding state for the superfluid state is not. 
As a matter of fact, the corresponding state for 
$(\sigma_{xx}^{H},\sigma_{xy}^{H})=(0,n)$ depends on the approaching
path. 
For our purpose, 
we are motivated to search for a critical flow without breaking the
parity and time-reversal symmetries, characterized by $\sigma_{xy,c} =0$. In
the context of two-parameter renormalization flow, this amounts to
finding the critical regions that follow a one-parameter 
(here it stands for $\sigma_{xx,c}$)
scaling. From Eq. (5), one has 
\begin{equation}
\sigma_{xy,c}=n^{2}\frac{\sigma_{xy,c}^{H}-n}{(\sigma_{xx,c}^{H})^{2}
+ (\sigma_{xy,c}^{H}-n)^{2}} + n = 0.
\end{equation}
The solution of the above equation leads to a flow line where 
the longitudinal and transverse quantum Hall 
conductivities follow a semi-circle relation:
\begin{equation}
(\sigma_{xx,c}^{H})^{2} 
+ (\sigma_{xy,c}^{H} - \frac{n}{2} )^{2} =(\frac{n}{2})^2 .
\end{equation}  
Summarized in Table I are the corresponding states in concord with the
one-parameter scaling flow. 
It is interesting to observe that an anyon superfluid
state without breaking the parity and time-reversal symmetries can be reached 
if the corresponding quantum Hall
conductivities following a semi-circle relation in the critical
region. From a general two-parameter scaling point of view,  
it is expected a general points at the $\sigma_{xx}^{H}-\sigma_{xy}^{H}$
plane will flow into the semi-circle critical regions. 
There exists three fixed points in the critical region: 
two stable ones at $(\sigma_{xx}^{H}, \sigma_{xy}^{H }) =
(0,0)$ and $(\infty, 0)$, corresponding  
to the insulator and superfluid states, respectively. 
The unstable one is located at $(\sigma_{xx}^{H \ast}, \sigma_{xy}^{H \ast})=
(n/2,n/2)$, corresponding a quantum normal critical point for the 
non-chirality superfluid-to-insulation 
transition point with a critical normal conductance $\sigma_{R}$.

\begin{table}
\caption{The corresponding states between the IQH system and
disordered anyon model phase diagrams, following the semi-circle 
relationship for conductivities of quantum Hall systems
that corresponds to non-chirality 
superfluid-to-insulator flow (with a 
universal critical conductance $\sigma_{R}$) for the disordered anyons.}
\begin{tabular}{cclcl}
\multicolumn{1}{c}{$\sigma_{R}$} &
\multicolumn{1}{c}{$(\sigma_{xx}^{H},\sigma_{xy}^{H})$} &
\multicolumn{1}{c}{State} &
\multicolumn{1}{c}{$(\sigma_{xx},\sigma_{xy})$} &
\multicolumn{1}{c}{State}
\\ \hline
 $n$ &$(0,0)$ & Insulating & $(0,0)$ & Insulating \\
 &$(0,n)$ & IQH plateau & $(\infty,0)$ & Superfluid \\
& $(\frac{n}{2},\frac{n}{2})$ & Critical; Normal &
$(n, 0)$ & Critical; Normal  \\
 $n^2$ &$(0,n-1)$ & IQH plateau & $(0,0)$ & Insulating \\
 &$(0,n)$ & IQH plateau & $(\infty,0)$ & Superfluid \\
  & $(\frac{1}{2}, n - \frac{1}{2})$ & Critical; Normal 
& $(n^2, 0)$ & Critical; Normal 
\end{tabular}
\end{table}

It is intriguing to note that the direct 
metal-insulator transition is exactly what  
Sheng and Weng [13] have identified 
based on their tight-binding lattice model study of the IQH effect. 
Apart from the case of $n=1$, it is 
different from the nearest-neighbor plateau transitions.
Our analysis shows that for this 
universality class, the critical fixed
points, $(\sigma^{H\ast}_{xx}, \sigma^{H\ast}_{xy}) = (n/2,n/2)$,
can be derived from the symmetry considerations of the
duality transformation. 

An important implication of the above mapping analysis is that the 
longitudinal conductance of the IQH system follows
a scaling form (using the self-dual 
transformation Eqs. (4-5) and the semi-circle relation) 
\begin{equation}
\frac{\sigma_{xx}^{H}}{\sigma_{xx}^{H\ast}} 
= \frac{2 (\sigma_{xx}/ \sigma_{R} )}{1+ (\sigma_{xx}/ \sigma_{R} )^2} .
\end{equation}
Remarkably, this is exactly the form 
that has been found empirically from numerical simulations [10,13].
 
It is gratifying to show that the direct transitions are intimately 
connected to the particle-vortex duality intrinsic in the
self-duality mapping. It becomes clear that the symmetry transformations used 
in the global phase diagram [4], 
however, are in general not in conformity to the 
self-duality. The self-duality between the particle and vortex 
yields non-trivial coupling and/or merging of the edge states.  

$Float$-$down$ $transition.-$With the increase of disorder, the  
IQH effect for the high Laudau band disappears one after another
[11]. As a result, 
the particle-vortex duality is broken.  Thus we need to consider the 
effect of disorder on the change of mapping relations.  
From the correlated RPA equations [17], it is worth noting
that for averaging over the disorder, the   
longitudinal and transverse conductivities appearing in Eqs (4-5) are treated
in the equal footing. The only part that needs to be reevaluated is the 
disconnected part that contributes to $\sigma_{R}$. In principle,  
the contribution depends on the
strength and the specific form of the 
disorder. However, one can consider the strong disorder case 
when all the extended states of the anyons are projected
onto the lowest Laudau level for composite fermions. 
In this limit, the vortices are pinned by disorder and bidding 
together as composite fermions, with charge $ne$. 
One can argue that the contribution of a charge $ne$ composite 
fermion with flux quantization $h/ne$ is $\sigma_{R}=n^2$.
This is equivalent to set
$\sigma_{R}$ equal to 1 in the unit of $e^{\ast 2}/h$ ($e^{\ast}=ne$), 
the value for the lowest Landau level.  

With $\sigma_{R}=n^2$, it is readily observable from Eqs. (4-5) that the
delocalization point is located at $\sigma_{xx}^{H\ast}=\frac{1}{2}$ and
$\sigma_{xy}^{H\ast}= n- \frac{1}{2}$, in concord with the float-up 
fixed points. Also listed in Table I are the 
corresponding states for this case.
In contrast to the direct transition scenario where
there exits a particle-vortex duality, 
here $(\sigma_{xx}^{H}, \sigma_{xy}^{H}) = (0, n-1)$ maps to the
insulating state $(\sigma_{xx}, \sigma_{xy}) = (0, 0)$.  
The float-down transitions can be viewed as follows: 
$(\sigma_{xx}^{H},    
\sigma_{xy}^{H}) = (0,n)$ maps to a superfluid state for $n$-anyons. 
Through a metal-to-insulator transition following a semi-circle law
\begin{equation}
(\sigma_{xx,c}^{H})^{2} 
+ (\sigma_{xy,c}^{H} - n +\frac{1}{2} )^{2} =(\frac{1}{2})^2 ,
\end{equation}  
which corresponds to a superfluid-to-insulator transition for 
$n-$anyons, 
it reaches 
$(\sigma_{xx}^{H},
\sigma_{xy}^{H}) = (0,n-1)$ that is an insulating state for $n-$anyons
as well as 
a superfluid state for $(n-1)-$anyons.  
This leads to successive multi-step transitions, in agreement with
the scenario provided by 
the global phase diagram for float-down transitions [4].
 
It is interesting to note that the projection of extended states onto
the lowest Laudau band for composite fermions
refers to a situation appropriate for the 
fractional quantum Hall systems. 
The symmetry transformations
employed in the global phase diagram are valid
for fractional quantum Hall transitions. 
Therefore, the float-down transition can be realized 
for system of composite fermions. In general, 
both direct and float-down transitions can happen in the IQH 
phase diagram. A detailed determination of the complicated 
phase boundaries requires
the reference of numerical simulations [9-13].  
It is attempting to argue that there exists two critical magnetic fields,
$B_{c1}$ and $B_{c2}$. When $B_{c1}$ is reached from low-$B$ side, the 
phase diagram shows  
direct transition, following a scaling behavior as given by Eq. (8) [10]. 
In the
region $B_{c1}< B < B_{c2}$, there exist regions for float-down transitions,
and the one-parameter scaling form of Eq. (8) is no longer valid. 
A detailed analysis of the IQH phase diagrams deserves further studies. 

$Semi$-$circle$ $relation.-$The 
above analysis can be extended to 
the general case of $n \rightarrow m$ plateau-plateau transitions. 
The critical region is defined by a semi-circle relation:
\begin{equation}
(\sigma_{xx,c}^{H})^{2} 
+ (\sigma_{xy,c}^{H} - \frac{n+m}{2} )^{2} =(\frac{n-m}{2})^2 .
\end{equation}  
Remarkably, this is exactly the form that Ruzin and co-workers [15]
proposed based on phenomenological arguments.  
Our analysis shows that this critical
region corresponds to a non-chirality superfluid-insulator transition  
governed by one parameter scaling. 
The success in extracting this phenomenological theory from the
mapping relations provides strong support for the correctness
and consistency of our
analysis. 

$Conclusions.-$The revisit of the mapping 
relations between the superfluid-to-insulator
transition and the metal-insulator transition of the IQH system reveals 
the existence of two types of universality classes 
in the IQH system: the direct transitions in the 
weak disorder limit, and the float-down transition in the strong
disorder limit. 
Our analysis is based on the physics intuition that  
the one-parameter scaling of the non-chirality 
superfluid-to-insulator transition corresponds to  
critical regions for 
the quantum Hall system. 
The resulting predictions provides
theoretical foundation for 
features observed in computer simulations based on the
lattice model [9-13], such as the the existence of plateau-insulator 
direct transition
universality class [11] and the 
universal scaling form of conductancei [10-11]. In addition,
the analysis shed important light into the microscopic origin of the 
phenomenological semi-circle relation [15] for the longitudinal and transverse
quantum Hall conductivities at the critical region.  

This work was supported by
Air Force Office of
Scientific Research Grant No. F49620-96-1-0211,
US Army Grant No. DAAE07-99-C-L065, and 
National Science Foundation Grant No. DMR-02-05328.

\end{document}